\documentclass[showpacs,preprintnumbers,amsmath,nofootinbib,amssymb,superscriptaddress]{revtex4}

\usepackage{caption}
\usepackage{subcaption}
\usepackage{amsmath,color}
\usepackage{amssymb}
\usepackage{bm}
\usepackage{graphicx}
\usepackage{multirow}

\usepackage{textcomp}

\allowdisplaybreaks[1]

\newcommand{\nn}{\nonumber \\}
\newcommand{\beq}{\begin{eqnarray}}
\newcommand{\eeq}{\end{eqnarray}}

\newcommand{\Slash}[1]{{\ooalign{\hfil/\hfil\crcr$#1$}}}

\usepackage{color}
\usepackage[makeroom]{cancel}
\usepackage[normalem]{ulem}
\definecolor{rbcolor}{rgb}{0.7,0.1,0}

\newcommand\rbout{\marginpar{\color{rbcolor}$\clubsuit$}\bgroup\markoverwith{\color{rbcolor}{\rule[0.4ex]{2pt}{0.8pt}}}\ULon}

\begin{document}


\title{\boldmath
A QCD analysis of  near-threshold quarkonium  leptoproduction \\at large photon virtualities
}

\author{Renaud Boussarie}
\author{Yoshitaka Hatta}
\affiliation{Physics Department, Brookhaven National Laboratory, Upton, New York 11973, USA}


\date{\today}

\begin{abstract}

We propose a novel approach to compute the cross section of near-threshold $J/\psi$ and $\Upsilon$ production  in electron-proton scattering at large photon virtualities $Q^2$ based on an operator product expansion. We show that the process can be used to extract the gluon part of the D-term gravitational form factor of the proton. At the subleading level, it is also sensitive to the trace anomaly effect of QCD.

\end{abstract}


%
%
%

\maketitle

%
%
%

\section{Introduction}

The exclusive photoproduction of $J/\psi$ near threshold has a long history \cite{Gittelman:1975ix,Camerini:1975cy,Bauer:1977iq}, starting almost immediately after the discovery of $J/\psi$. In the early days, it was one of the key measurements for reconfirming the existence of $J/\psi$,  as well as studying its basic properties such as the coupling to hadronic matter. Over the past two decades, theoretical interest in this reaction resurfaced every once in a while \cite{Kharzeev:1998bz,Brodsky:2000zc,Frankfurt:2002ka,Bosted:2008mn,Gryniuk:2016mpk,Hatta:2018ina,Xu:2019wso,Hatta:2019ocp} with different focuses, but it was not until recently that the subject draw a lot of  attention from the viewpoint of the nucleon structure. It has been suggested theoretically \cite{Kharzeev:1998bz,Hatta:2018ina} that the detailed behavior of the cross section near threshold is sensitive to the trace anomaly of QCD, hence it can shed light on the origin of the proton mass (see the related works in \cite{Hatta:2019lxo,Mamo:2019mka,Wang:2019mza}). This is one of the main motivations for the ongoing  experiments at Jefferson laboratory (JLab)  \cite{Ali:2019lzf,Joosten:2018gyo}. Moreover,  the National Academy of Science in the U.S. \cite{nas} has recently identified the proton mass problem  
 as one of the major scientific goals of the future Electron-Ion Collider (EIC)  \cite{Accardi:2012qut,Aidala:2020mzt}. The subject is also actively discussed in the context of the EIC in China \cite{priv}. It is then perfectly possible that the physics of  near-threshold production grows into an important sub-field in the EIC era. 

The existing theoretical approaches are roughly divided into two categories. The one that has been used since the 70s   
  \cite{Bauer:1977iq,Pumplin:1975fd,Barger:1975ng,Kharzeev:1998bz} 
 is to assume vector meson dominance (VMD) for the incoming photon. In this approach, the original problem $\gamma p \to J/\psi p'$ is reduced to forward scattering $pJ/\psi \to pJ/\psi$ which is more amenable to various theoretical tools.  A heavy quarkonium interacts with a hadron only via gluon exchanges. In the heavy quark mass limit, the interaction effectively becomes  local and the scattering amplitude is described by the moments of the gluon distribution  function. At the subleading level, it also depends on the gluon condensate in the proton $\langle p|F^2|p\rangle$ \cite{Luke:1992tm,Kharzeev:1998bz} which constitutes the major part of the QCD trace anomaly. 
 
 The second approach makes  use of `two-gluon form factors' $\langle p'|FF|p\rangle$ \cite{Brodsky:2000zc,Frankfurt:2002ka,Hatta:2018ina}. One of the distinctive features of  near-threshold quarkonium production is that the  momentum transfer $t=(p'-p)^2$ is large. However, in the VMD approach,  non-forwardness is trivialized even though the threshold value $\sqrt{-t_{th}}\approx 1.5$ GeV is comparable to the charm quark mass which is treated as the only hard scale of the problem.    In  \cite{Hatta:2018ina}, it has been shown via a holographic method   that the amplitude is proportional to the gravitational form factor $\langle p'|T^g_{\alpha\beta}|p\rangle$ where $T^g_{\alpha\beta}$ is the gluon part of the energy momentum tensor. Subsequently, the precise relation between the trace of $T^g_{\alpha\beta}$ and the gluon condensate operator  $F^2$ has been understood  \cite{Hatta:2018sqd,Tanaka:2018nae}. On the other hand, $\langle p'|T^g_{\alpha\beta}|p\rangle$ also contains the so-called D-term  which appears only in nonforward kinematics and which has attracted a considerable attention lately (see a recent review \cite{Polyakov:2018zvc} and references therein). The results of \cite{Hatta:2018ina,Hatta:2019lxo,Mamo:2019mka} suggest that near-threshold quarkonium production is a unique process that can directly access not only the gluon condensate, but also the {\it gluon} D-term. The latter aspect is quite complementary to the ongoing effort to extract the quark D-term from deeply virtual Compton scattering (DVCS) \cite{Burkert:2018bqq,Kumericki:2019ddg,Moutarde:2019tqa}.  
 Yet, holographic approaches are at best a model of QCD, and it remains to be seen to what extent the obtained  predictions are borne out in real QCD.  
 
Overall, the current theoretical status just described is  not totally  satisfactory. What is missing is a first-principle approach in QCD which can be systematically improved and compared to the data.  
The present work is a step toward this aim. Instead of photoproduction, we propose to study leptoproduction with large photon virtualities $q^2=-Q^2$.  So far, leptoproduction has received remarkably little attention in the context of threshold production, perhaps because  as $Q^2$ gets  larger, one is further away from the forward kinematics.   However,  the large-$Q^2$ region appears to be the cleanest setup from a perturbative QCD point of view. We in fact consider the limit $Q^2\gg M^2$ where $M$ is the quarkonium mass. 

This paper is organized as follows. In the next section we briefly review the kinematics of the reaction $\gamma p \to J/\psi p'$ near threshold. In section III, we lay out our strategy to compute the scattering amplitude based on operator product expansion (OPE). In section IV, we discuss the various two-gluon form factors involved. Then in section V, we numerically evaluate the cross section and study the impact of the D-term as well as the gluon condensate. Section VI is devoted to conclusions.

\section{Kinematics}

We shall be interested in the near-threshold production of a heavy quarkonium vector meson $H$ with mass $M$ in electron-proton scattering $ep \to e'\gamma^* p \to e'H p'$. We have in mind $H=J/\psi$ and $\Upsilon$. The center-of-mass energy of the virtual photon-proton subsystem at the threshold is
\beq
W^2_{th}=(q+p)^2=(m_N+M)^2, 
\eeq
where   $m_N=0.94$ GeV is the proton mass. Numerically, $W_{th}\approx 4.04$ GeV for $J/\psi$ and $W_{th}\approx 10.4$ GeV for $\Upsilon$. $q$ and $p$ are the virtual photon and proton momenta, respectively, with $Q^2=-q^2$ being the photon virtuality. 
Near the threshold, the Bjorken variable takes the form 
\beq
  x_B= \frac{Q^2}{2p\cdot q} \approx \frac{Q^2}{Q^2+M^2 + 2m_NM}. \label{bj}
\eeq
Eq.~(\ref{bj}) shows that, unlike the usual situation in DIS, $Q^2$ and $x_B$ are not independent variables. $x_B$ approaches unity as $Q^2$ goes to infinity. We also see that, somewhat counterintuitively,  there is no kinematical  restriction in $Q^2$. Threshold production can occur even when  $Q^2$ is arbitrarily large. 
Using the standard variables in DIS, $S_{ep}=(p+\ell)^2$, $y=p\cdot q/p\cdot \ell$ where $\ell^\mu$ is the incoming electron momentum, we can write 
\beq
W^2=y(S_{ep}-m_N^2)+m_N^2-Q^2. \label{w2}
\eeq
Small-$W$ does not necessarily imply small-$S_{ep}$ when $Q^2$ is large. In particular, the process can be  studied at the future EIC  \cite{Lomnitz:2018juf}.   

 Let $\vec{p}_{cm}$ and $\vec{k}_{cm}$ be the 3-momentum of the incoming proton and outgoing quarkonium, respectively,  in the center of mass frame of the $\gamma^*p$ subsystem.
\beq
p_{cm}^2&=&\frac{W^4-2W^2(m_N^2-Q^2)+(m_N^2+Q^2)^2}{4W^2}, \nonumber \\ 
k_{cm}^2&=&\frac{(W^2-(M+m_N)^2)(W^2-(M-m_N)^2)}{4W^2}.
\eeq
The momentum transfer is 
\beq
t&=& \left(\sqrt{p_{cm}^2+m_N^2} -\sqrt{k_{cm}^2+m_N^2}\right)^2 - \left(\vec{p}_{cm}+\vec{k}_{cm}\right)^2.
\eeq
At the threshold, $k_{cm}=0$ 
 so that 
\beq
t_{th}= 2m_N\left(m_N-\sqrt{p_{cm}^2+m_N^2}\right) = -\frac{m_N(M^2+Q^2)}{m_N+M}. \label{tmin}
\eeq
We see that $|t_{th}|$ is minimal in photoproduction $Q^2=0$ and monotonously increases with increasing $Q^2$. In the heavy-quark mass limit, $Q^2 \gg |t_{th}|$. 
Away from the threshold, $t$ takes a value in the range $|t_{max}|>|t|>|t_{min}|$ depending on the angle between $\vec{p}_{cm}$ and $\vec{k}_{cm}$.  
The differential cross section is given by 
\beq
\frac{d\sigma_{T/L}^{\gamma^*p}}{dt} = \frac{\alpha_{em}}{8 (W^2-m_N^2) W\, p_{cm}} \frac{1}{2}\sum_{spin}\left|\langle p'k|\epsilon^{T/L}_q\cdot J_{em}(0)|p\rangle\right|^2,
\eeq
 where $J_{em}^\mu = \sum_f e_f \bar{q}_f\gamma^\mu q_f$ is the electromagnetic current ($e_f$ being the charge in units of $|e|$) and $T/L$ refers to the transversely ($T$) or longitudinally ($L$) polarized virtual  photon. The factor $1/2$ is for averaging over proton helicities. The nontrivial dynamics of QCD is contained in the hadronic matrix element
\beq
\int d^4y e^{-iq\cdot y} \langle p'k| J^\nu_{em}(y)|p\rangle = (2\pi)^4\delta(k+p'-q-p) \langle p'k| J^\nu_{em}(0)|p\rangle.
\label{matt}
\eeq
Computing  (\ref{matt}) from first principles in QCD is a challenging task. Most of the previous theoretical works have focused on the photoproduction limit.  
In contrast, in this paper we shall investigate leptoproduction in the large $Q^2\gg M^2$ region.


\section{OPE at large $Q^2$}

In this section, we formulate our strategy to calculate the hadronic matrix element $\langle p'k|J_{em}|p\rangle$ near threshold at large $Q^2\gg M^2$. We have chosen to work in the high $Q^2$ region for reasons to become clear shortly. For definiteness, we consider $J/\psi$ production, but the case with $\Upsilon$ is completely analogous. In fact, our approach is better justified when $M\gg m_N$. Thus,  $\Upsilon$ production is more preferred from a theoretical point of view, though of course experimentally it is more challenging. 

Let us first mention that, if the center of mass energy is sufficiently high $s=W^2 \gg M^2,|t|$, the process is commonly called Deeply Virtual Meson Production (DVMP). The cross section is known to factorize in perturbative QCD in terms of the generalized parton distribution (GPD) and the meson distribution amplitude (DA)  \cite{Collins:1996fb}. Near the threshold, $s={\cal O}(M^2)$,  and $|t|$ is comparable to, or even exceeds $s$ depending on the value of $Q^2$, see (\ref{tmin}). Note,   however, that $s$ is small because of the cancellation $s=2p\cdot q-Q^2+\cdots$ and $x_B=Q^2/2p\cdot q$ stays close to unity. Moreover, $2p\cdot q\sim Q^2\gg |t|$ at least parametrically when $M\gg m_N$ (see (\ref{tmin})). This gives us some hope that a  perturbative approach is possible. 

Our basic argument is that near the threshold, the amplitude (\ref{matt}) is related to the following current-current correlator
\beq
\epsilon_\mu^*(k) 
i\!\int d^4x d^4y e^{ik\cdot x-iq\cdot y}
\langle p'|{\rm T}\{\bar{c}\gamma^\mu c(x) J^\nu_{em}(y)\}|p\rangle,  \label{argue}
\eeq
where $\epsilon_\mu(k)$ is the $J/\psi$ polarization vector. This matrix element is similar to the one that appears in  doubly virtual Compton scattering (DDVCS) $\gamma^{*}(q)p \to \gamma^*(k)p'$, or  timelike Compton scattering (TCS) in the special case $q^2=0$ (see, e.g., \cite{Berger:2001xd}). However, there is a  crucial difference. The DDVCS amplitude is given by the correlator $\langle J_{em}J_{em}\rangle$, and is dominated by the light quark degrees of freedom (light quark GPDs) except in the very small-$x_B$ region where it is dominated by gluons. In (\ref{argue}), on the other hand, one of the electromagnetic currents has been replaced by the charm quark current operator (bottom quark, in the case of $\Upsilon$ production). As a result, only the charm component of the other $J_{em}$ is relevant, and the matrix element becomes  primarily sensitive  to the gluonic content of the proton. 

That the $J/\psi$ production amplitude is related to a DDVCS-like (photon production) amplitude is intuitively reasonable, in view of the fact that in actual experiments, a $J/\psi$ and a timelike  photon with virtuality exactly at the $J/\psi$ mass are practically indistinguishable as they are probed via leptonic final states ($e^+e^-$ pairs). However,  
in DDVCS or TCS, the resonance region $k^2\approx M^2$ is usually avoided because the nonperturbative final state effect to produce the vector meson comes into play. 
As a function of $k^2$, the right hand side of (\ref{argue}) has a sharp resonance peak near the $J/\psi$ mass shell. 
Using the LSZ reduction formula, we can write \beq
ee_f \epsilon_\mu^*(k) \, i\!\int d^4x d^4y e^{ik\cdot x-iq\cdot y}
\langle p'|{\rm T}\{\bar{c}\gamma^\mu c(x) J^\nu_{em}(y)\}|p\rangle \approx \frac{ig_{\gamma J/\psi}}{k^2-M^2+iM\Gamma}  \int d^4y e^{-iq\cdot y} \langle p'k| J^\nu_{em}(y)|p\rangle, \label{lsz}
\eeq
where $e_f=2/3$ for the charm quark and $\Gamma$ is the total width of $J/\psi$. The decay constant is related to the  
electromagnetic width as 
\beq
\Gamma_{e^+e^-}= \frac{\alpha_{em} g_{\gamma J/\psi}^2}{3M^3}.
\eeq
Away from the very narrow peak (note that $\Gamma =93$ keV$\, \ll M=3.1$ GeV), the current correlator is expected to behave smoothly. We thus arrive at the relation 
\beq
 \int d^4y e^{-iq\cdot y} \langle p'k| J^\nu_{em}(y)|p\rangle =\beta \frac{ee_f M^2}{ g_{\gamma J/\psi}} \epsilon_\mu^*(k) \, i\!\int d^4x d^4y e^{ik\cdot x-iq\cdot y}
\langle p'|{\rm T}\{\bar{c}\gamma^\mu c(x) J^\nu_{em}(y)\}|p\rangle, \label{fin}
\eeq
where $\beta$ is a c-number of order unity which is not under control. It is understood that the right hand side is evaluated close to, but not too close to the $J/\psi$ mass shell $|k^2-M^2|\gg M\Gamma$. Our key observation is that in this off-mass-shell region, one can perform an operator product expansion (OPE) when $Q^2$ is large. 

Before doing so, a few additional remarks are in order. (i)
On general grounds, one expects corrections to (\ref{lsz}) from higher  resonances which the operator $\bar{c}\gamma^\mu c$ can excite.
However,  this effect will be  suppressed near threshold because, at fixed values of $s$ and $t$, only the resonances with mass smaller than $\sqrt{s}-m_N$ can be produced. 
There may also be  contributions from the deep Euclidean region $k^2<0$ if one considers a dispersion relation for the current correlator in $k^2$ similarly to \cite{Altarelli:1972sw,Pasquini:2001yy}.\footnote{We thank K. Tanaka for pointing this out.} Such an analysis may lead to a more precise evaluation of the quarkonium matrix element and help to determine the value of $\beta$. We leave this to future work. 
(ii) 
Our argument is similar in spirit to  the vector meson dominance (VMD) hypothesis. Note that this is different from the VMD assumption used  in many literature works on $J/\psi$ photoproduction mentioned in the introduction   \cite{Pumplin:1975fd,Barger:1975ng,Bauer:1977iq,Kharzeev:1998bz}. In these works, VMD has been applied to the incoming massless photon $\gamma \to J/\psi$. In photoproduction, this results in a significant mismatch between the initial and final virtualities $0\to M^2$. Here, in a sense, we apply VMD  in a reverse way to the outgoing $J/\psi \to \gamma^*$   (cf. \cite{Altarelli:1972sw,Lutz:2005yv,Gryniuk:2016mpk}). While the difference in virtualities partly remains, this has little impact on the overall kinematics of the reaction because $|k^2-M^2|\ll Q^2,|t|$.
(iii)
On the other hand, our approach is  different from the nonrelativistic (NR)QCD framework \cite{Bodwin:1994jh} which is commonly used for quarkonium production in hadronic collisions.  In NRQCD, the charm and anticharm quarks in the perturbative amplitude which couple to an external $J/\psi$ are both on-shell to leading order in the velocity expansion. However, in (\ref{fin}), $c$ and $\bar{c}$ are far off-shell with virtuality of the order of $Q^2$ (see below).  Moreover, (\ref{argue}) assumes that $J/\psi$ is produced only in a color-singlet state. This is reasonable because near the threshold, all the energy has to be used to create a $J/\psi$, and there is little phase space for extra gluon emissions.  
 
Let us now discuss the OPE. The current correlator on the right hand side of (\ref{fin}) can be written as 
\beq
&&i\int d^4 x d^4 y e^{ik\cdot x-iq\cdot y} \langle p'|{\rm T} \{ \bar{c}\gamma^\mu c(x)J_{em}^\nu(y)\}|p\rangle\nn
&& \qquad \approx e_f(2\pi)^4\delta^{(4)}(k+p'-q-p)\, i \int d^4r e^{ir\cdot \frac{k+q}{2}} \langle p'|{\rm T}\{\bar{c}\gamma^\mu c(r/2) \bar{c}\gamma^\nu c(-r/2)\}|p\rangle, \label{amp}
\eeq
where $r=x-y$.  The product of currents can be expanded if the relative distance $|r^\mu|$ is small, which is the case when the momentum $\frac{k+q}{2}$ is deeply spacelike. From (\ref{tmin}), near the threshold, 
\beq
t=(k-q)^2 = M^2+q^2-2k\cdot q \approx -\frac{m_N(M^2+Q^2)}{m_N+M}. \label{tt}
\eeq
Therefore,
\beq
(k+q)^2=M^2+q^2 +2k\cdot q \approx 2M^2-2Q^2 +\frac{m_N(M^2+Q^2)}{m_N+M}.
\eeq
This can be made arbitrarily negative by choosing $Q^2 \gg M^2$. We need to also make sure that the large momentum $Q$ does not `leak' into the proton vertex which in practice means $Q^2 \gg |t|$. Very close to  the threshold, this is satisfied if $M\gg m_N$. As one goes away (but not too far away) from the threshold,  the condition $Q^2\gg |t|$ is well  satisfied when $t\sim t_{min}$.\footnote{For example, if we set $W=4.4$ GeV  and $Q^2=100$ GeV$^2$, we find $|t_{max}|\approx 52$ GeV$^2$ and $|t_{min}| \approx 10$ GeV$^2$.} As we shall see in Section V,   $t\sim t_{min}$ is the most interesting region. 

However, for technical reasons the `symmetric' form (\ref{amp}) is not very convenient.
 Being a nonforward matrix element, it can be expressed in several different `frames'  
\beq
 i\int 
 d^4r e^{ir\cdot \frac{k+q}{2}} \langle p'|{\rm T}\{ \bar{c}\gamma^\mu c\left(\frac{r}{2}\right)\bar{c}\gamma^\nu c\left(-\frac{r}{2}\right)\}|p\rangle &=& i\int 
 d^4r e^{ir\cdot q }\langle p'|{\rm T}\{ \bar{c}\gamma^\mu c\left(0\right)\bar{c}\gamma^\nu c\left(-r\right)\}|p\rangle \nn 
 &=& i\int 
 d^4r e^{ir\cdot k} \langle p'|{\rm T}\{ \bar{c}\gamma^\mu c\left(r\right)\bar{c}\gamma^\nu c\left(0\right)\}|p\rangle.
 \label{mid}
 \eeq
The meaning of the OPE is different in different frames.  The final result must be the same, but this equivalence is often difficult to see. 
We shall return to this issue later.  For the moment  we find it most convenient to start with the middle expression of (\ref{mid}). 
We evaluate it as 
\beq
 i\int d^4r e^{ir\cdot q}\bar{c} \gamma^\mu c (0)\bar{c}\gamma^\nu c(-r) 
 &=& i \int d^4r e^{ir\cdot q} \bigl( \bar{c}(0) \gamma^\mu S(0,-r)\gamma^\nu c(-r) + \bar{c}(-r) \gamma^\nu S(-r,0)\gamma^\mu c(0)    \bigr) \nn 
&&   -i\int d^4r e^{ir\cdot q} {\rm Tr}[\gamma^\mu  S(0,-r)\gamma^\nu S(-r,0)] +\cdots, \label{so}
\eeq
where $S$ is the charm quark propagator in the presence of background gluon fields. 
 Since we work in the regime $Q^2\gg M^2$, $|r^\mu|$ is typically much smaller than $1/M$ and  the heavy quark mass $m_c \approx M/2$ can be neglected to first approximation. An important point of our approach is that 
  we shall expand (\ref{so}) in terms of {\it local} operators \cite{Watanabe:1981ue,Chen:1997rc}, instead of nonlocal light-cone correlators as is usually done in high energy scattering. Near the threshold, the role of light-cone directions appears to be less conspicuous. More importantly, the OPE with local operators is  well suited for our purpose of establishing a connection to the D-term which is the matrix element of the (local) energy momentum tensor operator. 
 
Consider the first line on the right hand side of (\ref{so}).  The lowest contribution comes  from the operator $\bar{c}\gamma^\mu\gamma_5c$, followed by higher dimensional operators such as $\bar{c}\gamma^\mu D^\nu c$, $\bar{c}\gamma^\mu F^{\alpha\beta} c$ and $\bar{c}\tilde{F}^{\alpha\beta}\gamma_\beta \gamma_5 c$. The (nonforward) matrix elements of these operators measure the intrinsic charm contents of the proton which are in general  believed to be tiny (see however, \cite{Brodsky:2000zc}).  In this paper we simply neglect all of them, although they can be straightforwardly restored  if need arises. 

We thus focus on the second line of (\ref{so}). Basically, we only  keep   dimension-4 purely gluonic   operators. This in particular includes the gluon part of the QCD energy momentum tensor 
 \beq
T_g^{\alpha\beta}=-F^{\alpha\rho}F^{\beta}_{\ \rho} + \frac{g^{\alpha\beta}}{4} F^{\mu\nu}F_{\mu\nu}, \label{energy}
\eeq
whose proton matrix element $\langle p'|T_g^{\alpha\beta}|p\rangle$ is what we are ultimately interested in.  
However, certain higher dimension operators are {\it a priori} not  suppressed. As in usual DIS or DVCS, the contribution of the leading twist  operator with Lorentz spin-$j$ is proportional to $(2q\cdot p/Q^2)^j \sim (1/x_B)^j$ and $x_B\sim 1$ for our problem.  The difficulty to sum over these higher spin operators with $j>2$  is  the reason why the local version of the OPE is not commonly used in DVCS. Here, however, we do not attempt to perform this summation.  Among the twist-two operators,  the energy momentum tensor $T_{\alpha\beta}^g$ with $j=2$ dominates in the sum when $Q^2$ is sufficiently large. The contributions from the other twist-2 operators with spin $j>2$ are relatively suppressed because their anomalous dimensions are nonvanishing. Admittedly, the rate of this suppression is slow, only logarithmic in $Q^2$, so a large leverage in $Q^2$ is needed to isolate the spin-2 component. While this may seem a difficult task, we point out that a very similar problem exists in the current strategy to extract the quark D-term from the DVCS data \cite{Burkert:2018bqq,Kumericki:2019ddg,Moutarde:2019tqa}. The subtraction constant in the dispersion relation between the real and imaginary parts of the Compton form factor, commonly denoted by $\Delta(t)$ \cite{Polyakov:2018zvc}, is given by the sum of infinitely many Gegenbauer coefficients $\Delta (t,Q^2) = d_1(t,Q^2)+d_3(t,Q^2)+\cdots$. In order to isolate the quark D-term $\propto d_1(t)$  which has the same anomalous dimension as the energy momentum tensor,  one needs a large leverage in $Q^2$ to disentangle different moments. Assuming that such an analysis is feasible at the future EIC, we expect that the same can be done for the gluon D-term.


 We shall work in Fock-Schwinger gauge $r_\mu A^\mu(r) =0$ for actual calculations. In this gauge, in the small-$r$ limit, the massless quark propagator in the background gluon field is given by, in $d=4-2\varepsilon$ dimensions (see for example, \cite{Shuryak:1981pi,Balitsky:1987bk})
\beq
S(r,0) &=&\frac{i\Gamma(d/2)}{2\pi^{d/2}} \frac{\Slash r}{(-r^2)^{d/2}} -\frac{ig\Gamma(1-\varepsilon)}{2^5\pi^{d/2}} \frac{r_\alpha F_{\mu\nu}(0)}{(-r^2)^{1-\varepsilon}}   (\gamma^\alpha\sigma^{\mu\nu}+\sigma^{\mu\nu}\gamma^\alpha)
  +i\frac{g^2\Gamma(-\varepsilon)(-r^2)^{\varepsilon}}{2^6\pi^{d/2}N_c} r^\alpha F^a_{\alpha\rho}F^{\rho}_{a\beta}(0)\gamma^\beta \nn
 &&-i\frac{g^2\Gamma(-\varepsilon)(-r^2)^\varepsilon }{3\cdot 2^6\pi^{d/2}} \left( \gamma^\alpha F_{\alpha\rho}F^{\rho}_{\ \beta}(0)r^\beta    + r^\alpha F_{\alpha\rho}F^{\rho}_{\ \beta}(0)\gamma ^\beta -\Slash r F_{\alpha\beta}F^{\alpha\beta}(0) +  \frac{2\varepsilon \Slash r}{r^2}r^\alpha F_{\alpha\rho}F^{\rho}_{\ \beta}(0)r^\beta\right)+\cdots, \label{ex}
\eeq
where $F^{\alpha\beta}=F_a^{\alpha\beta}t^a$ with ${\rm Tr}(t^at^b)=\delta^{ab}/2$ and our convention for the covariant derivative is $D^\mu=\partial^\mu+igA^\mu$. In the denominators, $r^2$ is short for  $r^2-i\epsilon$.    In (\ref{ex}), we have  kept only the terms which contribute to dimension-4 gluonic operators $FF$. At first sight, the dimension-3 operators of the form  $D_\alpha F_{\beta \gamma}$ are irrelevant because they are matrices in color space so when inserted in (\ref{so}), they either vanish after tracing over color indices or lead to operators with dimension-5 or larger. However, for the present problem, it turns out that they cannot be neglected. We shall discuss this later.  
Note that, since the Fock-Schwinger gauge breaks translational invariance, in general $S(r,0)\neq S(0,-r)$. However, for the terms listed in (\ref{ex}), the relation $S(r,0)= S(0,-r)$ actually holds.

In the second line of (\ref{so}), the unit operator can be neglected because we are computing the nonforward amplitude $\langle p'|1|p\rangle=0$.  Consider then the ${\cal O}(g^2 FF)$ terms in (\ref{ex}) which lead to a  logarithmically enhanced contribution as implied by the prefactor $\Gamma(-\varepsilon)$. 
Taking the trace of the $g^2FF$ terms in (\ref{ex}) in  color space, we find  
\beq
{\rm Tr}_{color} [S(r,0)] \sim i\frac{g^2\Gamma(-\varepsilon)(-r^2)^\epsilon}{3\cdot 2^5\pi^{d/2}}  \left(r^\alpha \gamma^\beta  \hat{T}_{\alpha\beta}^g(0)-\frac{\varepsilon}{2r^2}\Slash r \hat{T}^g_{\alpha\beta} r^\alpha r^\beta \right), \label{sec}
\eeq
where
\beq
\hat{T}_g^{\alpha\beta}\equiv -F_a^{\alpha\rho}F^{\beta}_{a\rho} + \frac{g^{\alpha\beta}}{d}F_a^{\mu\nu}F^a_{\mu\nu}, \label{trace}
\eeq
 is the traceless part of  the gluon part of the QCD energy momentum tensor. 
(\ref{sec}) explicitly shows that the logarithmic part is insensitive to the trace anomaly. Inserting the first term of (\ref{sec}) into the second line of (\ref{so}), we find 
\beq
&&-ig^2\frac{\Gamma(-\varepsilon)\Gamma(d/2)}{3\cdot 2^5\pi^{d/2+2}} \int d^d re^{ir\cdot q} \frac{r^\alpha r_\lambda}{(-r^2+i\epsilon)^{d/2-\varepsilon}} {\rm Tr}[\gamma^\mu \gamma^\lambda \gamma^\nu\gamma^\beta]  \hat{T}_{\alpha\beta}^g(0) \label{wt} \\ 
&& = \frac{\alpha_s}{12\pi} \left(\frac{1}{\varepsilon} - \ln (-q^2/\mu^2) +1+ \cdots \right)  \left(\frac{g^\alpha_\lambda}{q^2} -2(1+\varepsilon)\frac{q^\alpha q_\lambda}{(q^2)^2}\right) {\rm Tr}[\gamma^\mu \gamma^\lambda \gamma^\nu\gamma^\beta]  \hat{T}_{\alpha\beta}^g(0)\nn 
&& \to  -\frac{\alpha_s}{3\pi} \left(  \ln (-q^2/\mu^2) -1\right)  \left(\frac{g^\alpha_\lambda}{q^2} -2\frac{q^\alpha q_\lambda}{(q^2)^2}\right)(g^{\mu\lambda}g^{\nu\beta} -g^{\mu\nu}g^{\lambda\beta} +g^{\mu\beta}g^{\lambda\nu}) \hat{T}_{\alpha\beta}^g(0) -\frac{\alpha_s}{6\pi} \frac{q^\alpha q_\lambda}{(\ell^2)^2}     {\rm Tr}[\gamma^\mu \gamma^\lambda \gamma^\nu\gamma^\beta]  \hat{T}_{\alpha\beta}^g(0)  .
\nonumber
\eeq
Note that in the last step we have dropped the divergent piece $1/\varepsilon$. It can be absorbed into the renormalization of the twist-two operator $\hat{T}_c^{\mu\nu}\sim \bar{c}\gamma^{(\mu} D^{\nu)} c$ contained in the first line of (\ref{so}). The coefficient $\frac{\alpha_s}{3\pi}$ can be identified with the anomalous dimension $\gamma_{c\leftarrow g}$ of this operator.   As we already mentioned, we neglect the matrix element of (renormalized) $\hat{T}_c^{\mu\nu}$ so in practice the $1/\varepsilon$ simply disappear. 

The non-logarithmic terms in (\ref{wt}) combine with those from 
the second term of (\ref{sec}) and the square of the ${\cal O}(gF)$ term in (\ref{ex}). 
After a tedious but straightforward calculation, we arrive at the total $\alpha_s FF$ contribution
  \beq
{\cal A}^{\mu\nu} &\equiv & i\int d^4r e^{ir\cdot q}\bar{c} \gamma^\mu c (0)\bar{c}\gamma^\nu c(-r)  
\nn 
&& \approx  -\frac{\alpha_s(\mu_R)}{3\pi q^2} \Biggl[2 \ln (-q^2/\mu_R^2)  \left\{ \left(g^{\mu\alpha} -\frac{q^\mu q^\alpha}{q^2}\right)\left(g^{\nu\beta}-\frac{q^\nu q^\beta}{q^2}\right) + \frac{q^\alpha q^\beta}{q^2}\left(g^{\mu\nu}-\frac{q^\mu q^\nu}{q^2}\right) \right\} \hat{T}_{\alpha\beta}^g(0) \nn 
&& \qquad \qquad \qquad  -2\frac{q^\alpha q^\beta}{q^2} \left(g^{\mu\nu}-\frac{q^\mu q^\nu}{q^2}\right)  \hat{T}_{\alpha\beta}^g(0)    + 3\frac{q_\alpha q_\beta}{q^2} F^{\mu\alpha}F^{\nu\beta}(0) \Biggr] ,  \label{twist}
 \eeq
 where the operators are defined at the scale $\mu_R$. 
This is manifestly transverse with respect to $q$, i.e.  $q_\mu {\cal A}^{\mu\nu}={\cal A}^{\mu\nu}q_\nu=0$, as a consequence of the Ward-Takahashi identity. In Appendix \ref{sec:DIS}, we show that the forward matrix element of (\ref{twist}) reproduces the 1-loop coefficient functions of the DIS structure functions.
However,  (\ref{twist}) has an obvious problem. The tensor ${\cal A}^{\mu\nu}$ is transverse with respect to $q^\mu$ and $q^\nu$, but this is because we have started with the middle expression in (\ref{amp}). In the present problem, gauge invariance rather implies $k^{\mu}{\cal A}_{\mu\nu}={\cal A}_{\mu\nu}q^\nu=0$.   Actually, problems of this kind typically arise in off-forward kinematics. It is known that ensuring the electromagnetic gauge invariance of DVCS amplitudes is a highly nontrivial issue \cite{Anikin:2000em,Belitsky:2005qn}. The leading order (leading twist) result does not fully satisfy the WT identity, and one has to include higher twist corrections to restore it. In the context of OPE, this amounts to including   operators with total derivatives \cite{Braun:2011dg}. 
In Appendix \ref{sec:derivatives}, we demonstrate that the dimension-3 operators $D_\alpha F_{\beta\gamma}$ which were neglected in (\ref{so}) indeed give rise to total derivative operators. This calculation suggests that a complete treatment of the problem requires the inclusion of dimension-5 and even dimension-6 operators in the expansion (\ref{so}), which is beyond the scope of this work. Here instead, we suggest an {\it ad hoc} solution of the problem. In (\ref{twist}), we set $q^2=-\mu_R^2$ to eliminate the logarithmic terms. In the remainder terms, we implement the following minimal modifications\footnote{There is an ambiguity when replacing $q^2$ with $q\cdot k=q^2-q\cdot \Delta$, since    $q^2\to \left(\frac{q+k}{2}\right)^2= q^2-q\cdot \Delta+\Delta^2/4$ seems to be an equally good choice (cf. (\ref{amp})) . However, the difference is subleading because $q^2\approx 2q\cdot k\gg \Delta^2$ in the present kinematics, see (\ref{tt}). This ambiguity can only be resolved by including the dimension-6 operator  $\partial^2T_{\alpha\beta}$.} to make $A^{\mu\nu}$ transverse with respect to $k^\mu$ and $q^\nu$, and symmetric in $q$ and $k$ 
\beq
{\cal A}^{\mu\nu} \to -\frac{\alpha_s}{3\pi (q\cdot k)^2} \left[ -2q^\alpha k^\beta  \left(g^{\mu\nu}-\frac{q^\mu k^\nu}{q\cdot k}\right)\hat{T}_{\alpha\beta}^g+
3k_\alpha q_\beta F^{\mu\alpha}F^{\nu\beta}\right], \label{pa}
\eeq
where the coupling and the operators are evaluated at the scale ${\cal O}(Q^2)$. 
In the `leading-twist' approximation, one can further simplify (see (\ref{weyl}))
\beq
-F^{\mu\alpha}F^{\nu\beta} \approx \frac{1}{2}(g^{\mu\nu}\hat{T}^{\alpha\beta}_g -g^{\mu\beta}\hat{T}_g^{\alpha\nu} -g^{\alpha\nu}\hat{T}_g^{\mu\beta}+g^{\alpha\beta}\hat{T}_g^{\mu\nu}). \label{cons}
\eeq
Actually, since we are neglecting the twist-2 operators with spin $j>2$, it is not entirely consistent to include anything beyond (\ref{cons}) as it  corresponds  to  twist-4 effects. Still, for phenomenological purpose it may be interesting to include at least the trace part of $T_g^{\alpha\beta}$ in order to assess the impact of the trace anomaly.

\section{Two-gluon form factors}

In order to  compute the actual cross section, we need to parametrize the non-forward matrix element of two-gluon operators in (\ref{pa}) in terms of form factors. First we have the gravitational form factors at our disposal \cite{Ji:1996ek}
\beq
\langle p'|T^{\mu\nu}_{g}|p\rangle = \bar{u}(p')\Bigl[ A_{g}\gamma^{(\mu}P^{\nu)} + B_{g}\frac{P^{(\mu}i\sigma^{\nu)\alpha}\Delta_\alpha}{2m_N} + D_{g}\frac{\Delta^\mu\Delta^\nu-g^{\mu\nu}\Delta^2}{4m_N} + \bar{C}_{g}m_Ng^{\mu\nu} \Bigr] u(p),  \label{grav}
\eeq
where $\Delta^\mu = p'^\mu - p^\mu$,   $P^\mu\equiv\frac{p^\mu+p'^\mu}{2}$ and $A^{(\mu}B^{\nu)} \equiv (A^\mu B^\nu + A^\nu B^\mu)/2$. All four form factors are functions of $t=\Delta^2$ and the renormalization scale $\mu_R$ in the $\overline{\rm MS}$ scheme.  $D_g$ is  the gluon part of the D-term form factor which we are mainly interested in. (In the literature often the notation $C_g=D_g/4$ is often used.) The $\bar{C}_g$ form factor is related to the trace anomaly \cite{Hatta:2018sqd}.  
The traceless part reads  
\beq
\langle p'|\hat{T}^{\mu\nu}_{g}|p\rangle = \bar{u}(p')\left[ A_{g}\gamma^{(\mu}P^{\nu)} + B_{g}\frac{P^{(\mu}i\sigma^{\nu)\alpha}\Delta_\alpha}{2m_N} + \frac{D_{g}}{4m_N} \left(\Delta^\mu\Delta^\nu-\frac{g^{\mu\nu}}{d}\Delta^2\right) -\frac{m_Ng^{\mu\nu}}{d}\left(A_g + \frac{\Delta^2}{4m_N^2}B_g\right)  \right] u(p). \label{traceless}
\eeq
Next consider the two gluon operator with four open indices 
\beq
\langle p'| -F_a^{\mu\alpha}F_a^{\nu\beta}|p\rangle. \label{mat}
\eeq
Its most general parametrization consistent with parity, hermiticity and time-reversal symmetry is\footnote{Terms which contain the antisymmetric tensor $\epsilon^{\mu\alpha\rho\lambda}$ are not independent. For example, the following identity holds
\beq
i\bar{u}' \epsilon^{\mu\alpha\rho\lambda}\gamma_5\gamma_\rho\bar{P}_\lambda u = m_N \bar{u}' i\sigma^{\mu\alpha}u + \frac{1}{2}\bar{u}'(\Delta^\mu \gamma^\alpha-\Delta^\alpha\gamma^\mu)u.
\eeq
 }

\beq
\langle p'| -F_a^{\mu\alpha}F_a^{\nu\beta}|p\rangle &=& \frac{A}{2}\bar{u}(p')(g^{\mu\nu}\gamma^{(\alpha} P^{\beta)}- g^{\mu\beta}\gamma^{(\alpha} P^{\nu)} -g^{\alpha\nu}\gamma^{(\mu} P^{\beta)} + g^{\alpha\beta}\gamma^{(\mu} P^{\nu)}) u(p)  \nonumber \\ 
&&+ \frac{B}{4m_N}\bar{u}(p')\left(g^{\mu\nu}i\sigma^{(\alpha\lambda}\Delta_\lambda P^{\beta)}-g^{\mu\beta}i\sigma^{(\alpha\lambda}\Delta_\lambda P^{\nu)}-g^{\alpha\nu}i\sigma^{(\mu\lambda}\Delta_\lambda P^{\beta)} +g^{\alpha\beta} i\sigma^{(\mu\lambda}\Delta_\lambda P^{\nu)}\right) u(p) \nonumber \\
&& + \frac{D}{8m_N}\bar{u}(p') \Bigl( g^{\mu\nu} \Delta^\alpha\Delta^\beta-g^{\alpha\nu}\Delta^\mu \Delta^\beta+ g^{\alpha\beta} \Delta^\mu\Delta^\nu-g^{\mu\beta}\Delta^\alpha\Delta^\nu \Bigr) u(p) \nonumber \\ 
&&+\frac{W}{3}m_N \bar{u}(p')(g^{\mu\nu}g^{\alpha\beta}-g^{\mu\beta}g^{\alpha\nu})u(p)  \nonumber \\ 
&& + \frac{X}{2m_N^2} \bar{u}(p') \left( (\gamma^\mu\Delta^\alpha - \gamma^\alpha\Delta^\mu)(P^\nu \Delta^\beta -P^\beta \Delta^\nu) +   (P^\mu \Delta^\alpha-P^\alpha\Delta^\mu)  (\gamma^\nu\Delta^\beta - \gamma^\beta\Delta^\nu) \right)  u(p) \nonumber \\ 
&& + \frac{Y}{m^3_N} \bar{u}(p')    (P^\mu \Delta^\alpha-P^\alpha\Delta^\mu)(P^\nu \Delta^\beta -P^\beta \Delta^\nu)   u(p) \nonumber \\ 
&& + \frac{Z}{4m_N} \bar{u}(p') \left(i\sigma^{\mu\alpha}(P^\nu\Delta^\beta-P^\beta\Delta^\nu)+i\sigma^{\nu\beta}(P^\mu \Delta^\alpha-P^\alpha\Delta^\mu)\right)u(p),  \label{para}
\eeq
The seven form factors can be partly constrained by requiring consistency with (\ref{grav}). 
Contracting  the indices $\alpha\beta$ in (\ref{para}), we get 
\beq
\langle p'| -F_a^{\mu\alpha}F^{\nu}_{a\alpha}|p\rangle &=& \frac{A}{2}\bar{u}(p')(m_N g^{\mu\nu}+2\gamma^{(\mu} P^{\nu)}) u(p) + \frac{B}{4m_N}\bar{u}(p')\left(g^{\mu\nu}i\sigma^{\alpha\lambda}\Delta_\lambda P_{\alpha}+2i\sigma^{(\mu\lambda}\Delta_\lambda P^{\nu)}\right) u(p) \nonumber \\
&& + \frac{D}{8m_N} \bar{u}(p') \Bigl( g^{\mu\nu} \Delta^2 +2 \Delta^\mu\Delta^\nu \Bigr) u(p) +Wm_N g^{\mu\nu} \bar{u}(p')u(p)  \nonumber \\ 
&& +\frac{X}{m_N^2} \bar{u}(p') (P^{(\mu}\gamma^{\nu)}\Delta^2 +m_N\Delta^\mu \Delta^\nu) u(p)  + \frac{Y}{m^3_N}\bar{u}(p')(P^\mu P^\nu \Delta^2+P^2\Delta^\mu \Delta^\nu)u(p) \nonumber \\ 
&& + \frac{Z}{4m_N}\bar{u}(p') \left( 2i\sigma^{(\mu\alpha}P^{\nu)}\Delta_\alpha+\Delta^\mu \Delta^\nu \right)u(p)
\nn
&=& A_g\bar{u}(p') \gamma^{(\mu} P^{\nu)} u(p) +\left(\frac{A }{2}+W+ \frac{2D_g+D}{8m_N^2}\Delta^2  + \frac{B\Delta^2}{8m_N^2}\right) g^{\mu\nu}m_N\bar{u}(p')u(p)  \nonumber \\ 
&&+ \frac{B_g}{2m_N}\bar{u}(p') i\sigma^{(\mu\lambda}\Delta_\lambda P^{\nu)} u(p)  + \frac{D_g}{4m_N} \bar{u}(p') \Bigl( \Delta^\mu\Delta^\nu -g^{\mu\nu}\Delta^2\Bigr) u(p), \label{na}
\eeq
where in the second equality we used the following relations which can easily be obtained by term-by-term comparison:
\beq
A+\frac{\Delta^2}{m_N^2}(X+Y)=A_g,\\
B+Z-\frac{\Delta^2}{m_N^2}Y =B_g,\\
 D+4X+4Y+\left(Z-\frac{\Delta^2}{m_N^2}Y\right)=D_g.
\eeq
 We see that only two linear combinations of $X,Y,Z$ enter  these relations.

By comparing the coefficients of $g^{\mu\nu}$, one should be able to obtain another relation between $W$ and $\bar{C}_g$. However, this is nontrivial due to the presence of the QCD trace anomaly.   
If one naively contracts the indices $\mu\nu$ in (\ref{na}) and computes the matrix element of $T_g^{\mu\nu}$ by forming the linear combination (\ref{energy}), one ends up with a wrong relation $\bar{C}_g= -A_g/4$ (in the forward limit) and $W$ is undetermined. The problem is intimately tied to operator  renormalization. In dimensional regularization, the following innocent-looking relation does not hold
\beq
g_{\mu\nu} (F^{\mu\alpha}F^{\nu}_{\ \ \alpha}) \neq F^2.
\eeq
Namely, operator renormalization and trace operation do not commute. The correct way to proceed is to  write 
\beq
 -F_a^{\mu\alpha}F^{\nu}_{a\alpha} = T_g^{\mu\nu}-\frac{g^{\mu\nu}}{4}F^2,
 \eeq
on the left hand side of (\ref{na}) and sum over the indices $\mu\nu$ using (\ref{grav}) and (\ref{na}). 
 This gives 
\beq
\langle p'|F^2|p\rangle= \left[-2A_g-4W +4\bar{C}_g- \left( \frac{3D_g}{2} +\frac{B_g}{2}-4(X+Y) -Z+\frac{\Delta^2}{m_N^2}Y\right) \frac{\Delta^2}{m_N^2}\right]m_N\bar{u}(p')u(p). \label{la1}
\eeq
 On the other hand, the matrix element $\langle p'|F^2|p\rangle$ has to be carefully evaluated in a chosen regularization scheme \cite{Hatta:2018sqd,Tanaka:2018nae} (see also \cite{Rodini:2020pis}). In dimensional regularization, it is given by a linear combination of the  gravitational form factors, see Eq.~(13) of Ref.~\cite{Hatta:2019lxo}
 \beq
\langle p'|F^2|p\rangle= \left[ K_g(A_g+4\bar{C}_g) + K_q (A_q+4\bar{C}_q) +(K_gB_g+K_qB_q -3K_gD_g-3K_qD_q )\frac{\Delta^2}{4m_N^2}\right] m_N\bar{u}(p')u(p), \label{la2}
\eeq
  where the quark gravitational form factors $A_q,B_q,D_q,\bar{C}_q$ are defined analogously to (\ref{grav}) for the quark part of the energy momentum tensor. The coefficients $K_{q,g}$ are defined in \cite{Hatta:2019lxo} and can be evaluated, in principle, to arbitrary order in perturbation theory. At the moment, the three-loop results are available \cite{Hatta:2018sqd,Tanaka:2018nae}. They depend on the number of flavors and the renormalization scale via the QCD coupling $\alpha_s(\mu_R)$. (\ref{la1}) and (\ref{la2}) give a complicated relation between $X,Y,Z,W$ and the quark and gluon gravitational form factors. In the forward limit $t=0$ it somewhat simplifies  and we find
\beq
4W(0)= 4\bar{C}_g(0)(1-K_g+K_q) -(2+K_g-K_q)A_g(0) -K_q,
\eeq
where we used $A_q(0)+A_g(0)=1$ and $\bar{C}_q(t)+\bar{C}_g(t)=0$. The relation to the parameter $b$ often used in the literature \cite{Ji:1994av}  is 
\beq
b\equiv \frac{\langle p|(1+\gamma_m)\sum m_f \bar{q}_f q_f|p\rangle}{2m_N^2}, \qquad   1-b =  \frac{\langle p|\frac{\beta(g)}{2g}F^2|p\rangle}{2m_N^2} = \frac{\beta}{2g} \left(  (A_g(0)+4\bar{C}_g(0))(K_g-K_q)+K_q\right), \label{b}
\eeq
where $\gamma_m$ is the mass anomalous dimension. $b$ is the partition of the trace anomaly into the quark and gluon condensates. It is scheme and scale dependent.

\section{Numerical results}

In this section we show numerical results for the differential cross section based on the formula (\ref{pa}). We do not intend to perform a complete calculation which is anyway not possible at the moment as it requires the detailed knowledge of  all seven form factors $A(t),B(t),..$. On the other hand, some information about the gravitational form factors $A_g,B_g,D_g,\bar{C}_g$ is already available in the literature. Based on this, we consider two interesting cases which allow us to make a quantitative prediction. Case 1: We use the `leading-twist' approximation (\ref{cons}) and  keep only the traceless part of the energy momentum tensor (\ref{traceless}). As explained in Section III, in doing so we assume that the contributions from the  twist-2 operators with spin $j>2$ can be either neglected or separated out by using a large leverage in $Q^2$. Case 2:  We evaluate the full two-gluon operators (\ref{pa}) including  the trace part of the energy momentum tensor. While this is not a consistent approximation (because we keep the twist-4 effect and neglect the twist-2, spin-$j>2$ contributions), it is an instructive  exercise to  assess the impact of the trace anomaly.  
In both cases, we set $B_g=0$ following the suggestion from  lattice QCD  (see, e.g., \cite{Alexandrou:2017oeh})  that this form factor is numerically small. In Case 2, we also set $X=Y=Z=0$ as we know nothing about these form factors. On the other hand, the $W$ form factor is related to the trace anomaly and will be given full consideration. 

We use the following parametrization of the gravitational form factors   
\beq
A_g(t) &=& \frac{A_g(0)}{(1-t/m_A^2)^2} , \qquad  A_q(t) = \frac{1-A_g(0)}{(1-t/m_A^2)^2}, \nn
 D_g(t) &=& \frac{D_g(0)}{(1-t/m_C^2)^3}, \qquad 
\bar{C}_g(t) = \frac{\bar{C}_g(0)}{(1-t/m_A^2)^2},
\eeq
 The tripole form of the D-term is motivated by the quark counting rule  \cite{Tanaka:2018wea}.
Since the form factors are evaluated at a large scale $\mu_R^2=Q^2$, to first approximation we can use the asymptotic results
\beq
A_q(0)\approx \frac{n_f}{4C_F + n_f}, \qquad  A_g(0) \approx \frac{4C_F}{4C_F + n_f}, \qquad  D_q(0) \approx \frac{n_f}{4C_F}D_g(0),
\eeq
with $C_F=\frac{N_c^2-1}{2N_c}$ and $n_f=3$ represents the number of light flavors in the proton. The value $D_g(0)$ is our main interest and should be determined by future experiments. Here, for the sake of demonstration, we use the results of a recent lattice simulation $D_g(0)=-7.2$ ($C_g(0)=-1.8$) with $m_A=1.13$ GeV and $m_C=0.76$ GeV at $\mu_R=2$ GeV \cite{Shanahan:2018nnv}. 
(We neglect the scale dependence of these parameters.)
On the other hand $\bar{C}_g$ at zero momentum transfer is related to the QCD trace anomaly \cite{Hatta:2018sqd}.  Asymptotically $\mu_R^2\to \infty$,
\beq
\bar{C}_g(0) \approx \frac{1}{4} \left(\frac{n_f}{4C_F+n_f}+\frac{2n_f}{3\beta_0}\right)-\frac{1}{4}\left(\frac{2n_f}{3\beta_0}+1\right)\frac{b}{1+\gamma_m}, \label{cbar}
\eeq
where $b$ is introduced in (\ref{b}). To one-loop, $\beta_0=11N_c/3-2n_f/3$ and $\gamma_m= \frac{3C_F\alpha_s}{2\pi}$.     
A more precise expression can be found in \cite{Hatta:2018sqd,Tanaka:2018nae}.

Under these assumptions,  (\ref{la1}) and (\ref{la2}) reduce to a simple formula 
\beq
4W(t)=  4\bar{C}_g(t)(1-K_g+K_q) -\left(2+K_g+\frac{1-A_g(0)}{A_g(0)}K_q\right)A_g(t)   \nn 
+ 3D_g(t) \left(K_g +\frac{n_f}{4C_F}K_q-2\right) \frac{\Delta^2}{4m_N^2}.
\eeq
For simplicity, we use the one-loop result $K_{q,g}$
\beq
K_g= \left({\cal C}_{gF}-\frac{{\cal C}_{gm}}{{\cal C}_{qm}}{\cal C}_{qF}\right)^{-1}, \qquad K_q = -\frac{ {\cal C}_{gm}}{{\cal C}_{qm}}K_g,
\eeq
where ($\alpha_s=\alpha_s(Q^2)$)
\beq
{\cal C}_{qm} = 1+\frac{C_F}{3\pi} \alpha_s, \quad
{\cal C}_{qF} = \frac{n_f}{12\pi} \alpha_s, \\
{\cal C}_{gm} = \frac{7C_F}{6\pi}\alpha_s,  \quad
{\cal C}_{gF} = -\frac{11N_c}{24\pi} \alpha_s.
\eeq
See \cite{Tanaka:2018nae} for the three-loop result.
 Finally, the square of the prefactor in (\ref{argue}) is evaluated as (including the factor $e_f$ from (\ref{amp}))  
\beq
\frac{e_c^4e^2M^4}{g^2_{\gamma J/\psi}} \approx 24.6, 
\label{naive}
\eeq
where we set $e_c=2/3$, $M=3.1$ GeV, 
$\Gamma_{e^+e^-}=5.55$ keV and   $\alpha_{em}=e^2/4\pi=1/137$. The corresponding value for $\Upsilon$ is 
\beq
\frac{e_b^4e^2M^4}{g^2_{\gamma \Upsilon}} \approx 19.4, 
\label{naive2}
\eeq
 where $e_b=-1/3$, $M=9.46$ GeV and $\Gamma_{e^+e^-}=1.34$ keV. 
The parameter $\beta$ should be determined by fitting the data (for example the total cross section at some value of $W$) for each quarkonium species. In the numerical results below we set  $|\beta|=1$.

\begin{figure}
\begin{subfigure}{.5\textwidth}
  \includegraphics[width=0.9\linewidth]{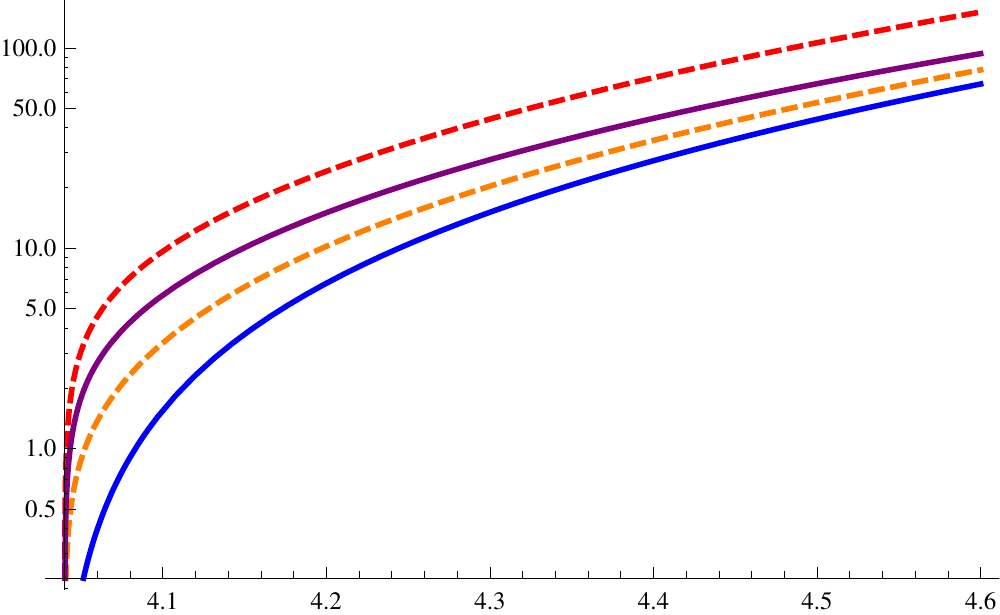}
  \caption{Total cross section (in fb) as a function of $W$ (in GeV).}
  \label{fig:sub1}
\end{subfigure}%
\begin{subfigure}{.5\textwidth}
  \includegraphics[width=0.9\linewidth]{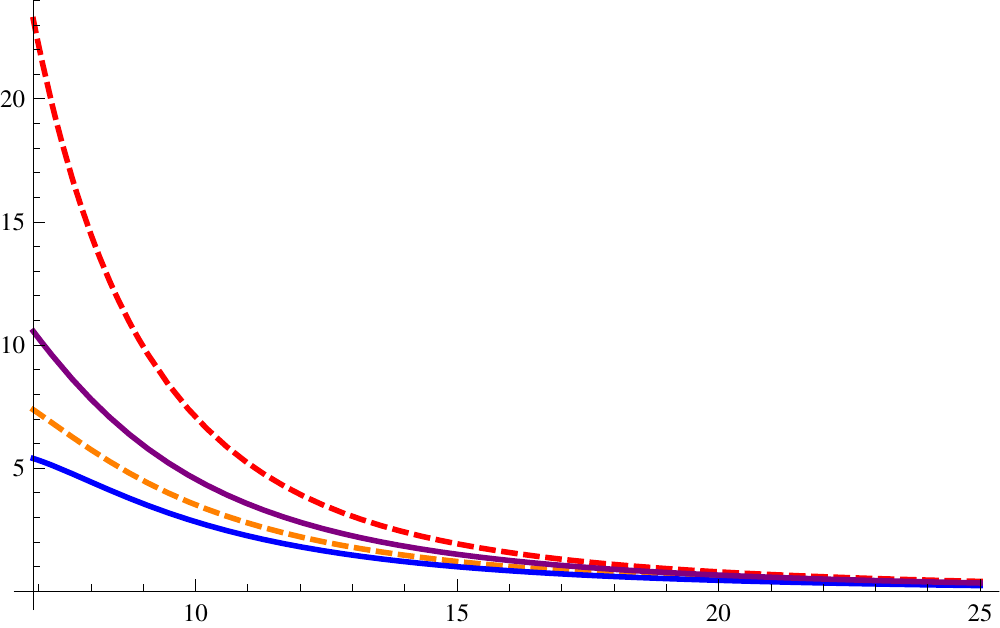}
  \caption{Differential cross section (in fb/GeV$^2$) at $W=4.4$ GeV as a function of $|t|$ (in GeV$^2$).}
  \label{fig:sub2}
\end{subfigure}
\caption{
$J/\psi$ total and differential cross sections at $Q^2=64$ GeV$^2$.  The upper and lower dashed curves correspond to Case 1 with $D_g=0$ and $D_g=-7.2$, respectively.   The upper and lower solid curves correspond to Case 2, $D_g=-7.2$, with $b=1$ and $b=0$, respectively.
}
\label{jpsi}
\end{figure}

In Fig.~\ref{jpsi}, we show the total and differential cross sections for $J/\psi$ at $Q=8$ GeV,  $\alpha_s(Q)=0.2$. The latter is evaluated at $W=4.4$ GeV. In both plots, the upper and lower dashed curves correspond to Case 1 with 
 $D_g=0$ and $D_g=-7.2$, respectively. We see a dramatic impact of the gluon D-term.\footnote{Remember that we neglected the RG evolution of $D_g$ from $\mu_R=2$ GeV to $Q=8$ GeV. The value $|D_g(0)|$ at the scale $Q$ will be smaller than 7.2 so the actual difference between the two dashed curves is expected to be smaller.}  A negative (positive) D-term tends to shrink  (enhance) the differential cross section. The same tendency has been observed in \cite{Hatta:2018ina} in the case of photoproduction $Q^2=0$. 
The upper and lower solid curves correspond to Case 2 with $b=1$ (zero gluon condensate) and $b=0$ (zero quark condensate), respectively.
 We see that the dependence on the parameter $b$ is significant. We also see that the gluon condensate tends to reduce the cross section, which is actually opposite to what was found in \cite{Hatta:2018ina}. It is not clear to us whether this is due to the fact that different  processes were considered  (photoproduction vs.  leptoproduction), or perhaps due to  the deficiency of the model used in  \cite{Hatta:2018ina}.

\begin{figure}
\begin{subfigure}{.5\textwidth}
  \includegraphics[width=0.9\linewidth]{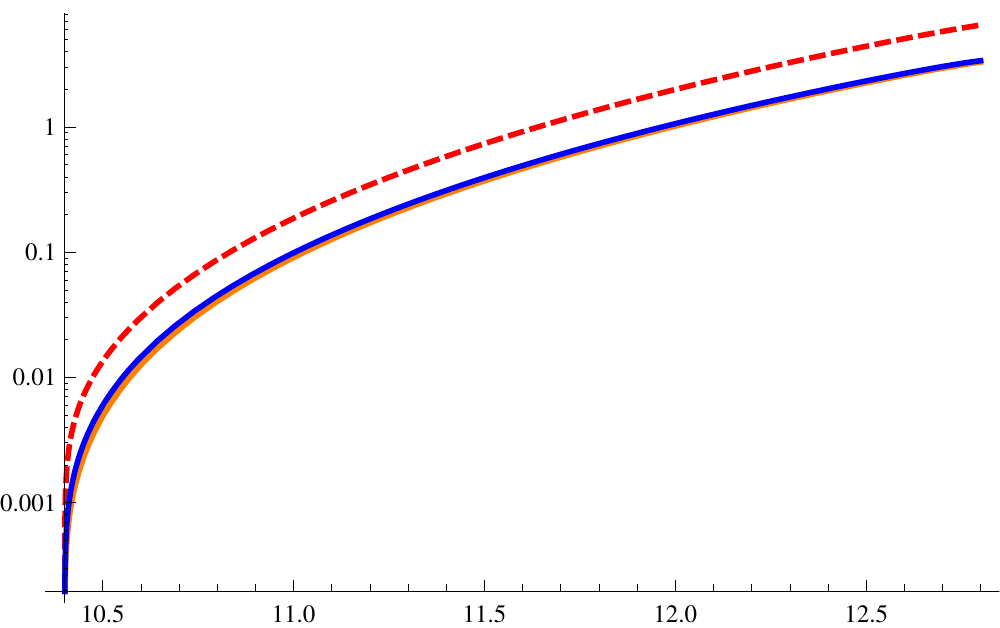}
  \caption{Total cross section (in fb) as a function of $W$ (in GeV).}
  \label{fig:sub1}
\end{subfigure}%
\begin{subfigure}{.5\textwidth}
  \includegraphics[width=0.9\linewidth]{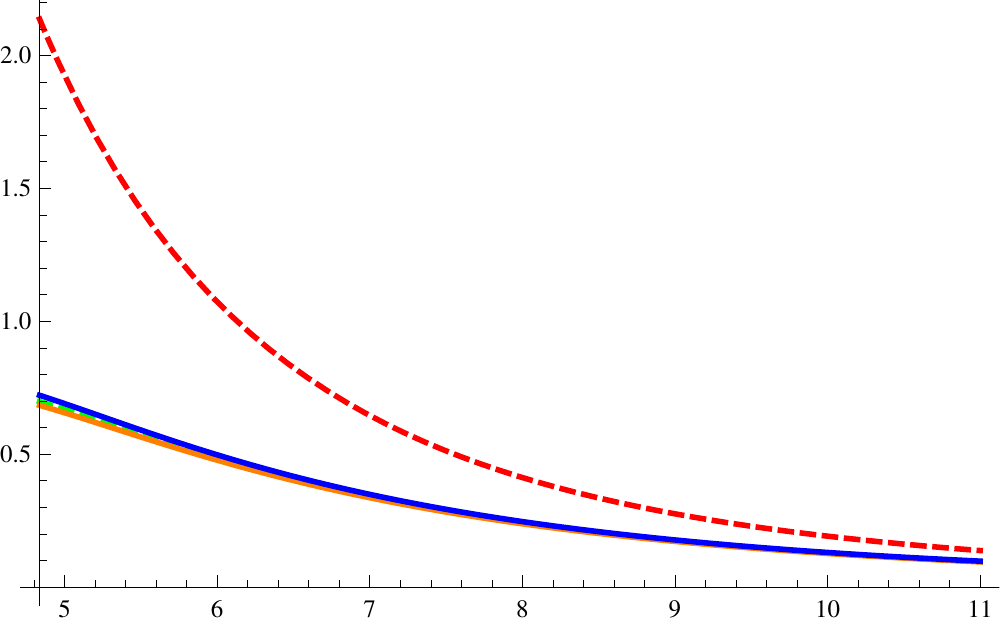}
  \caption{Differential cross section (in fb/GeV$^2$) at $W=12.5$ GeV as a function of $|t|$ (in GeV$^2$).}
  \label{fig:sub2}
\end{subfigure}
\caption{
$\Upsilon$ total and differential cross sections at $Q^2=18^2$ GeV$^2$. See the caption of Fig.~\ref{jpsi} for the  explanation of each curve.
}
\label{upsilon}
\end{figure}

Next, in Fig.~\ref{upsilon} we show the result for $\Upsilon$ at $Q=18$ GeV, $\alpha_s(Q)=0.16$.  Near the threshold ($W_{th}=10.4$ GeV), the cross section becomes very small. In the right panel we selected a somewhat large value  $W=12.5$ GeV considering the realistic luminosity of EIC. Again we see a large effect of the D-term. However, the impact of the trace anomaly and the split between $b=1$ and $b=0$ are barely visible.

In photoproduction, the $J/\psi$ total cross section  is about 1 nb at $W=4.5$ GeV \cite{Ali:2019lzf}. 
In leptoproduction, we see that the cross section is several orders of magnitude smaller. 
 For $\Upsilon$ production, there is another 2 orders of magnitude suppression. Thus, near-threshold leptoproduction is a luminosity-hungry observable.  Moreover, as  explained in Section III, one needs a large leverage in $Q^2$ to extract the D-term. 
Given these requirements, we think that the best place to test our proposal is $J/\psi$ (and possibly also $\Upsilon$) production in the high luminosity mode of EIC.

\section{Conclusion}

In this paper we have proposed a novel  strategy to compute the cross section of near-threshold quarkonium production at large momentum transfer. Compared to  photoproduction, near-threshold leptoproduction has so far attracted much less attention due to the lack of strong phenomenological motivations. We have demonstrated that the process is useful for probing the gluon D-term, quite complementary to the ongoing effort to extract the quark D-term in DVCS.   
The possible impact of the D-term on the differential cross section $d\sigma/dt$ has been already pointed out in the case of photoproduction using holography  \cite{Hatta:2018ina,Mamo:2019mka}. In leptoproduction at large $Q^2\gg M^2$, the problem can be studied within the perturbative framework. Moreover, at the subleading level the cross section is also sensitive to the value of the parameter $b$ defined in Eq.~(\ref{b}) which characterizes the structure of the QCD trace anomaly. The proposed measurements require high luminosity and a large leverage in $Q^2$. The only machine that can deliver these requirements is the future EIC. 

Our analysis in this paper is only the first step and there are a number of directions for future research.  
In particular, it is interesting to see if a similar approach can be applied to photoproduction using the heavy quark mass as a hard scale.  On the phenomenological side, the contribution from the Bethe-Heitler process $e^-\to e^-\gamma^*$ needs to be investigated  as $J/\psi$ and $\Upsilon$ are  reconstructed from lepton pairs in actual experiments.  The seven form factors introduced in (\ref{para}) can be calculated in lattice QCD. The values of $t$ considered in this paper are rather large, and the extrapolation to the forward limit is a serious challenge. Lattice  calculations of these form factors will be very valuable in this respect.

\section*{Acknowledgements}

 We are grateful to Jian-Wei Qiu and Kazuhiro Tanaka for discussions and critical comments, and Peter Schweitzer for correspondence.  
This work is supported by the U.S. Department of Energy, Office of
Science, Office of Nuclear Physics, under contract No. DE- SC0012704,
and in part by Laboratory Directed Research and Development (LDRD)
funds from Brookhaven Science Associates and by grant no. 2019/33/B/ST2/02588 of the National Science Center in Poland.

\appendix 
\section{DIS coefficient  functions}\label{sec:DIS}

 As a consistency check, let us compute the forward matrix element of (\ref{twist}) in a single proton state  and keep only the twist-2 contribution.   In this approximation,  we can write 
 \beq
 \langle p|\hat{T}_{g}^{\alpha\beta}|p\rangle = 2A_{g} p^\alpha p^\beta, 
\eeq
 where $A_g$  is the  fraction of the proton momentum  carried by gluons. 
 We then decompose the operator $F^{\mu\alpha}F^{\nu\beta}$ as\footnote{This decomposition is mathematically identical to that of the Riemann tensor in general relativity. The trace part $T_{\alpha\beta}$ is an analog of the Ricci tensor which represents the matter content and   $C^{\mu\alpha\nu\beta}$ is an analog of the Weyl tensor which represents the  gravity degrees of freedom.} 
\beq
F^{\mu\alpha}F^{\nu\beta}&=&\frac{1}{d-2}(g^{\mu\nu}F^{\lambda \alpha}F_{\lambda}^{\ \beta} -g^{\mu\beta}F^{\lambda\alpha}F_{\lambda}^{\ \nu} -g^{\alpha\nu}F^{\lambda\mu}F_\lambda^{\ \beta}+g^{\alpha\beta}F^{\lambda\mu}F_\lambda^{\  \nu})\nn 
&& \qquad +\frac{1}{(d-2)(d-1)}(g^{\mu\beta}g^{\nu\alpha}-g^{\mu\nu}g^{\alpha\beta})g_{\rho\sigma}F^{\rho\lambda}F^{\sigma}_{\ \lambda} + C^{\mu\alpha\nu\beta},
\label{weyl}
\eeq
and extract the energy momentum tensor component $T^{\alpha\beta}_g\sim -F^{\alpha\lambda}F^{\beta}_{\ \lambda}$. 
The remainder tensor $C^{\mu\alpha\nu\beta}$ has the same symmetry as $F^{\mu\alpha}F^{\nu\beta}$ except that it is traceless with respect to any pair of indices. Its forward matrix element  vanishes.  We thus have 
 \beq
 \langle p|-F^{\mu\alpha}F^{\nu\beta}|p\rangle \approx  A_g (g^{\mu\nu}p^\alpha p^\beta -g^{\mu\beta}p^\alpha p^\nu -g^{\alpha\nu} p^\mu p^\beta + g^{\alpha\beta}p^\mu p^\nu).
 \eeq
 This gives 
 \beq
 && i\int d^4r e^{ir\cdot q}\langle p|\bar{c} \gamma^\mu c (0)\bar{c}\gamma^\nu c(-r) |p\rangle \nn 
&& \approx 2A_g \frac{\alpha_s}{\pi (q^2)^2 }\Biggl[ \frac{2}{3}  (p\cdot q)^2 \left(g^{\mu\nu}-\frac{q^\mu q^\nu}{q^2}\right)     -\left(\frac{2}{3}\ln \frac{-q^2}{\mu_R^2} -\frac{1}{2}\right) q_\alpha q_\beta(g^{\mu\nu}p^\alpha p^\beta -g^{\mu\beta}p^\alpha p^\nu -g^{\nu\alpha}p^\mu p^\beta + g^{\alpha\beta}p^\mu p^\nu ) \Biggr].
 \eeq
From this one can read off the known one-loop coefficient functions for the DIS structure functions
\beq
 C_{2,2}^G= \left( \frac{2}{3}\ln \frac{-q^2}{\mu_R^2} -\frac{1}{2}\right)\frac{\alpha_s}{4\pi}  , \qquad C_{L,2}^G= \frac{2}{3} \frac{\alpha_s}{4\pi},
\eeq
in the notation of \cite{Larin:1996wd}.

\section{Total derivative operators}\label{sec:derivatives}

In this appendix, we give an example of how operators with total derivatives enter the calculation. We return to (\ref{ex}) and include dimension-3 operators $D_\alpha F_{\beta\gamma}$ which have been previously neglected
\beq
S(r,0) \sim
 \frac{4g}{3\cdot 2^6 \pi^{d/2}}(-r^2)^\varepsilon \Biggl[ \Gamma(-\varepsilon) D^\rho F_{\rho\lambda}\gamma^\lambda +\frac{1}{r^2} \left(\Slash r r^\lambda  D^\rho F_{\rho\lambda} +r^\alpha r^\beta D_\beta F_{\alpha\lambda}\gamma^\lambda +3ir^\alpha r^\beta D_\alpha \tilde{F}_{\beta\lambda}\gamma^\lambda \gamma_5\right)\Biggr], \label{non}
\eeq
\beq
S(0,-r) \sim 
 \frac{4g}{3\cdot 2^6 \pi^{d/2}}(-r^2)^\varepsilon \Biggl[ \Gamma(-\varepsilon) D^\rho F_{\rho\lambda}\gamma^\lambda +\frac{1}{r^2} \left(\Slash r r^\lambda  D^\rho F_{\rho\lambda} +r^\alpha r^\beta D_\beta F_{\alpha\lambda}\gamma^\lambda - 3ir^\alpha r^\beta D_\alpha \tilde{F}_{\beta\lambda}\gamma^\lambda \gamma_5\right)\Biggr]. \label{non}
\eeq
This can be derived following \cite{Shuryak:1981pi,Balitsky:1987bk}. Note that the last term proportional to $\gamma_5$ breaks the naive relation $S(r,0)=S(0,-r)$.  
Let us focus only on the singular term $\propto \Gamma(-\varepsilon)$  which is sufficient to demonstrate our  point. 
Its contribution to the current correlator is 
 \beq
&&\frac{\alpha_s \Gamma[1-\varepsilon]}{3\cdot 2^5 \pi^{d-1}}\int d^d r e^{-ir\cdot q} \frac{r_\alpha}{(-r^2)^{1-2\varepsilon}}\Gamma(-\varepsilon) \tilde{F}^{\alpha\lambda} D^\sigma F_{\sigma\rho}{\rm Tr}[\gamma^\mu \gamma_\lambda \gamma_5\gamma^\nu\gamma^\rho] \nn 
&&=i \frac{\alpha_s (4\pi)^\varepsilon}{3\pi} \frac{\Gamma(2+\varepsilon)\Gamma[1-\varepsilon]}{\Gamma(1-2\varepsilon)} \frac{q_\alpha (-q^2)^{-\varepsilon}}{(q^2)^2} \epsilon^{\mu\lambda\nu\rho}\tilde{F}^{\alpha}_{\ \lambda} D^\sigma F_{\sigma\rho}\Gamma(-\varepsilon) \nn 
&&= - i \frac{\alpha_s}{3\pi(\ell^2)^2} \left(\frac{1}{\varepsilon}+1-\ln \frac{(-q^2)}{\mu_R^2 }\right) (q^\mu \partial_\rho T_g^{\rho\nu} -q^\nu \partial_\rho T_g^{\rho\mu}  - F^{\mu\nu} q^\rho D^\sigma F_{\sigma \rho} ),  \label{sign}
 \eeq 
where we used the identity $\partial_\nu T_g^{\mu\nu} = F_{\nu}^{\ \mu}D_\alpha F^{\alpha\nu}$. Note that this is antisymmetric in $\mu$ and $\nu$. 
The $1/\varepsilon$ divergence in (\ref{sign}) is absorbed into the renormalization of the operator 
\beq
\bar{c}\gamma^\mu g\tilde{F}^{\alpha}_{\ \lambda}\gamma_\lambda \gamma_5 \gamma^\nu c - (\mu \leftrightarrow \nu) \sim  \epsilon^{\mu\lambda\nu\rho}\bar{c}\gamma_\rho g \tilde{F}^{\alpha}_{\ \lambda} c,
\eeq
which comes from the first line of (\ref{so}). The remaining terms contain total derivative operators. 
In the nonforward matrix element, the derivative operator is replaced by the momentum transfer  $\langle p'|\partial_\rho T_g^{\rho \nu}|p\rangle= i\Delta_\rho \langle p'|T_g^{\rho\nu}|p\rangle$ where $\Delta_\rho = q_\rho -k_\rho$. This is how, in principle, total derivative operators from higher dimensional terms can restore the WT identity through the addition of $\Delta$ corrections.
However, (\ref{sign}) is not sufficient to make  the logarithmic terms in (\ref{twist}) transverse with respect to $k^\mu$. For that, we would need  operators like $\partial^\nu T_g^{\mu\beta}$ and $g^{\mu\nu}\partial_\alpha T_g^{\alpha\beta}$.  We presume that the missing terms come from the dimension-5 and dimension-6 operators in the expansion of $S(r,0)$. We leave this to a future work.

\end{document}